\documentclass[12pt,reqno]{article}
\usepackage{latexsym,amsmath,amsthm,amsfonts,amssymb,amscd,endnotes}
\usepackage[dvips]{graphicx,psfrag}

\newcommand{\p}{{\bf p}}
\newcommand{\R}{{\bf r}}
\newcommand{\dR}{\dot{\R}}

\newcommand{\V}{{\bf v}}

\newtheorem{theo}{Theorem}

\newtheorem{lemma}{Lemma}

\title{\bf Some insights from  total collapse}
\author{S\'ergio B. Volchan}

\begin{document}

\date{}
\maketitle

\vskip .5cm
\begin{flushright}
\end{flushright}

\begin{abstract}

We discuss the Sundman-Weierstrass theorem of total collapse in its historical context. This remarkable
and relatively  simple result, a type of stability criterion, is at the crossroads of some interesting
developments in the gravitational Newtonian $N$-body problem. We use it as motivation to explore 
the connections to such important concepts as integrability, singularities and typicality in order 
to gain some insight on the transition from a predominantly quantitative to a novel qualitative  
approach to dynamical problems that took place at the end of the 19th century. 

\end{abstract}



\bigskip

\renewcommand{\thesection}{\Roman{section}}
\section{Introduction}

Celestial mechanics is one of the treasures of physics. With a rich and fascinating 
history,~\endnote{See, for instance, C. M. Linton, {\em From Eudoxus to Einstein. A History
of Mathematical Astronomy} (Cambridge University Press, NY, 2004) and also M. Gutzwiller, 
``Moon-Earth-Sun: the oldest three-body problem'', 
Reviews of Modern Physics, {\bf 70} (2), 589-639 (1998).} it
had a pivotal role in the very creation of  modern science. After all, it was Hooke's question on the 
two-body problem and  Halley's encouragement (and financial resources) that eventually led Newton 
to publish the {\em Principia}, a watershed.~\endnote{V. I. Arnold, {\em Huygens \& Barrow, Newton \&
Hooke} (Birkh\"auser, Basel, 1990), Ch. 1.} Conversely, celestial
mechanics was the testing ground {\em par excellence} for the new mechanics, and its triumphs in 
explaining  a wealth of phenomena were decisive in the acceptance of the Newtonian synthesis and  
the ``clockwork  universe''.

From the beginning, special attention was paid to the $N$-body problem,
the study of the motion of $N$ massive bodies under mutual gravitational forces, taken as a
reliable model of the the solar system. The list of scientists that worked on it, starting with Newton 
himself, makes a veritable hall of  fame of mathematics and physics; also, a host of concepts and 
methods were created  which are now part of the vast heritage of modern  mathematical-physics: 
the calculus, complex  variables, the theory of errors and statistics,  differential equations,  
perturbation theory, potential theory,  numerical methods, analytical mechanics, 
just to cite a few.

It is true that around the second-half of the 19th century, the  mainstream of physics
was less interested in what might have seemed a noble but old-fashioned subject, and 
attention turned to the exciting new areas of thermodynamics and electromagnetism
(and later, relativity, quantum theory and astrophysics). However, the field continued to be 
pursued by first-rate mathematicians and astronomers, who didn't lose sight
of  its enduring relevance. Henri Poincar\'e, one of the masters of the field, put it 
with utmost clarity when he observed that, besides the compilation of ephemeris, ``the ultimate goal 
of celestial mechanics is to resolve the great problem of determining if Newton's laws alone explain all 
astronomical phenomena''.~\endnote{See D. L. Goroff, ``Henri Poincar\'e and the birth of chaos theory: an
introduction to the english translation of {\em Les M\'ethodes Nouvelles de la
M\'ecanique C\'eleste}'' in  {\em
New Methods of Celestial Mechanics (History of Modern Physics)},  H. Poincar\'e and D. L. Goroff (Ed.), 
Vol. 1 (AIP Press, 1992), p. I17.} In this regard, recall that one of the first breakthroughs of 
Einstein's general relativity theory was an explanation of the discrepancy in the observed precession of 
Mercury's orbit,  a long-standing open problem in classical celestial mechanics.

The arrival  of the space-age (and race) and the advances in computer technology in the middle of the 
20th century, sparked a renewed interest in celestial mechanics, in connection to problems of spacecraft 
navigation and solar systems dynamics. New exciting ideas appeared, particularly from 
the American and soviet schools: new approaches to perturbation theory  (as in KAM theory), the challenges 
of chaotic behavior (whose existence was first glimpsed  in Poincar\'e's work on the three-body problem), new tools
from nonlinear dynamics, all leading to a rethinking of the implications to the long-time behavior of the solar system 
(the old stability problem).

This trend continues today,  with an intense cross-fertilization between the highly 
abstract tools from nonlinear dynamical systems and  sophisticated  computer simulations, spurred by 
applications to astrodynamics.~\endnote{For a review, see J. E. Marsden and 
S. D. Ross, ``New methods in celestial mechanics and mission design'', Bull. AMS, {\bf 43}, 43-73 
(2006)} Recent highlights are the unraveling of the interplanetary superhighway and the discovery of 
new solutions of the three-body problem.~\endnote{See, respectively, 
D. L. Smith, ``Next Exit 0.5 Million Kilometers'',  Engineering \& Science, {\bf LXV} (4), 6-15, (2002) 
and R. Montgomery, ``A New Solution to the Three-Body Problem'', Notices of the AMS, {\bf 48} (5), 471-481 (2001).} These 
developments should be a sobering alert against a certain ``anti-classical'' bias of modern physics education, which may 
impart to students the impression that classical physics is completed  (or stagnant), as if the only interesting and important 
questions lie in the quantum realm.

In this paper we present the Sundman-Weierstrass theorem of total collapse in its historical context. This 
relatively simple result, which is a kind of stability condition for the $N$-body problem,
deserves to be better known. Its surprisingly simple proof, using a little calculus and
linear algebra, can be profitably presented in a  general mechanics course. Besides, it can be used as 
motivation to address many issues of historical and conceptual interest. Our main aim is to stimulate the reader's curiosity 
to explore this vast field and to experiment a bit of what had been aptly described as ``the sheer joy of celestial  
mechanics''.~\endnote{From N. Grossman, {\em The Sheer Joy of Celestial Mechanics} (Birkh\"auser, Boston, 1996).}

The paper is structured as follows. Section II describes the mathematical formulation
of the $N$-body problem and the question of its  integrability. In section III we see how celestial mechanics
reached a crisis near the end of the 19th century, which  revolved around  the changing notion of
a solution of the $N$-body problem.  Section IV deals with the intriguing issue of singularities and the need
to circumvent them in the quest for exact solutions for the three-body problem; this leads directly to
the total collapse theorem which is also proved there for the general case of the $N$ bodies. 
We conclude by arguing that the events described were part of a major conceptual change that took place in dynamics, leading
from a predominantly quantitative to a new  qualitative approach, the impacts of which are still being assimilated. 

\renewcommand{\thesection}{\Roman{section}}
\section{The N-body problem and its integrability}

A fundamental problem of celestial mechanics is {\em the $N$-body problem}, which consists in 
applying Newton's laws of mechanics and his law of universal gravitation to an isolated system of $N$ point 
masses moving in three-dimensional space.  With respect to an inertial 
reference frame and  disregarding the influence of other bodies, we have for the position
$\R_j$ of the $j$th body,
\begin{equation}
\label{motion}
\displaystyle m_j  {\ddot \R}_j =  \nabla_{\R_j} U ({\bf x})=  
\Big(\frac{\partial U}{\partial x_{j1}}({\bf x}), 
 \frac{\partial U}{\partial x_{j2}}({\bf x}), \frac{\partial U}{\partial x_{j3}}({\bf x})\Big),
\end{equation}
plus the initial conditions: at the initial time $t_0$, we are given the positions 
 $\R_j(t_0) \neq \R_j(t_0)$,  $i\neq j$, and velocities  e $\V_j(t_0)=\dot{\R}_j(t_0)$. Here,  
${\bf x}=(\R_1, \ldots, \R_N)$  is the system's 
configuration and $U$ is the  gravitational potential energy 
(using the sign convention used in celestial mechanics):
\begin{equation}
\label{pot}
 U({\bf x}) \equiv \sum_{1\leq j <k \leq N} \frac{G m_i m_j}{\|\R_k -\R_j\|}, 
\end{equation}
where  $\| \cdot \|$ is the usual euclidean distance. 
Despite being an idealization (almost a caricature) of real gravitational
systems, this model is nonetheless very successful and still in current use.~\endnote{Additional credibility to
the model came from Newton's proof that the gravitational force outside  a spherically symmetric body 
is the same as if all mass were concentrated in its center. However, the oblate shape is  more realistic
and can have important effects.}  

As is well known, Newton's equations have an equivalent Hamiltonian version, namely, for $j=1, \ldots, N$:
\begin{equation}
\label{ham}
\begin{cases}
\dR_j = \nabla_{\p_j} H & \\
\displaystyle{\dot \p}_j = -\nabla_{\R_j} H
\end{cases}
\end{equation}
where $\p_j= m_j \V_j$, $H=T-U$ is the Hamiltonian function and with given initial condition 
$({\bf x}(t_0), {\bf p}(t_0))$ belonging to the system's
{\em phase-space}  $(\mathbb R^{3N}-\Delta)  \times  \mathbb R^{3N}$.~\endnote{Observe the exclusion of 
configurations belonging to the {\em singular set}  $\displaystyle\Delta  = \cup_{1 \leq i < j\leq N} 
\Delta_{ij}$, where $\displaystyle \Delta_{ij}=\{{\bf x}=(\R_1, \cdots, \R_N) \in \mathbb R^{3N}: \; 
\R_i=\R_j\}$.} 

Mathematically, ~\eqref{ham} is a system of $6N$ non-linear first-order differential equations; 
together with the initial data it defines an initial value problem. Faced with such 
a problem, the first task of a modern mathematician would be to prove the existence and uniqueness of solutions. 
To a physicist or  engineer, however, ``this is like leaving a restaurant without having eaten anything, 
but having paid the cover charge for the privilege of reading the menu''.~\endnote{M. Bunge, {\em Chasing
Reality. Strife over Realism} (University of Toronto Press, Toronto, 2006), pp. 151.} Here we have a classical 
instance of the clash of styles of physicists and  mathematicians, an endless source of 
witticisms.~\endnote{For example, an anecdote tells of a lecture delivered by mathematician Mark Kac at Caltech, to which Feynman attended. After the lecture Feynman got
up and announced: ``If all mathematics disappeared, it would set physics back by precisely
one week.'' To which Kac immediately replied: ``Precisely the week in which God created
the world.''}

Though there is a genuine difference of concerns, it is more fruitful to see these 
approaches as complementary. So, if a physicist thinks his model gives a reasonable description of 
the phenomenon at hand,  he is  confident (maybe more than the mathematician) the formalism should be
sound, and proceeds to get the consequences of it. On the other hand, an existence and uniqueness
theorem (EUT) is not a mere academic exercise;~\endnote{By the way,  such a theorem,
first due to Cauchy, was  distilled  from some numerical methods used in practice by astronomers; see 
A. Wintner, {\em Analytical Foundations of Celestial Mechanics} (Princeton University Press, 1964), p. 143.}  
besides  vindicating the soundness of the formalism,  it is the basis of the influential principle of Laplacian 
(or ``mechanical'') determinism: given the initial conditions, the future and the past  are completely determined.  
Interestingly, the EUT is usually a purely local result in that it only asserts the existence of solutions for a small 
time interval. What kind of ``catastrophe'' could prevent the existence of global solutions? We'll come
back to this in section IV.

Meanwhile, we observe that traditionally the main interest regarding the $N$-body problem was in its
{\em integration}, that is, in finding its {\em solution} giving  the positions of the bodies  
as ``explicit'' functions of time. It turns out that integrability is a thorny issue; to begin with,
what is meant by  such vague terms  as  ``explicit function'', ``closed formula'' or 
``analytic expression''?

Though there are different notions of integrability available, historically it usually referred
to  ``integration  by quadratures''  (also known as ``reduction''): a solution is to be 
found by performing a finite number of algebraic operations, integrations and  inversions, over 
a class of known ``elementary functions''.  Key to the method is finding so called 
``first integrals'' (or constants of motion), i.e., functions of positions and momenta which are 
constant along the solution. As the  level set of such a function (corresponding to given initial data) 
defines a hyper-surface in phase-space to which the solution curve belongs, the  dimension of the 
problem is  thereby reduced by one. Thus, if  enough independent first integrals can be found
the problem eventually becomes one-dimensional, and could be solved  by ``simple'' integration.

 The $N$-body problem has ten classical independent first integrals, namely: the total energy and
the components of, respectively, the total linear momentum, the total angular momentum and 
the center-of-mass.~\endnote{As usual, we take the center-of-mass reference frame 
so that  ${\bf R}_{cm}={\mathbf P}= \mathbf 0$.}  These  correspond
to the conservation laws associated to the Galilean invariance of classical 
mechanics.~\endnote{They were known to Lagrange as of 1772, in the context of the three-body problem,
who also obtained an additional reduction by  ``elimination of time''.}  So, in principle, the system 
could be reduced to a  $(6N-11)$-dimensional one.

The two-body problem is integrable by quadratures as  $6N-11=1$. However, already for the 
three-body problem the classical integrals are not enough, reduction leading to a $7$-dimensional
system.~\endnote{Jacobi later discovered an additional  ``reduction of the nodes''.} This led to
a conundrum clearly summarized by Wintner:~\endnote{See ref. 11.}
\begin{quote}
``When John and James Bernoulli, Clairaut, D'Alambert, D. Bernoulli, and Lambert, Euler and, finally,
Lagrange applied the principles of Newton to the various problems of celestial and terrestrial mechanics,
they had to face an awkward situation. For, on the one hand, it was almost axiomatic that a dynamical 
problem is `solved' only if it is reduced to quadratures (and successive differentiations and
eliminations); while on the other hand, the most urgent problems were almost never reducible by 
quadratures.''
\end{quote}
 Of course, people had used other means, if not to exactly solve, at least 
to extract useful information from  the $N$-body problem; for instance, looking for special
solutions (say, with some symmetry) or  using perturbation and numerical methods. However, and 
probably under the spell of the success with the two-body problem, there was a widespread conviction 
that sooner or later  the ``geometers'' would find the missing integrals, 
which  would lay bare the secrets of the three-body problem.~\endnote{See J. Laskar, ``La stabilit\'e 
du syst\`eme solaire'', in {\em Chaos et d\'eterminisme}, edited by A. D. Dalmedico, J. -L. Chabert and 
K. Chemla (\'Editions du Seuil, Paris, 1992), pp. 171-211.} The
search for new  integrals began to look like a kind of holy grail saga.

\renewcommand{\thesection}{\Roman{section}}
\section{The Oscar prize}

Though integration by quadratures was made mathematically rigorous by Sophus Lie,~\endnote{See V. I. Arnold,
V. V. Kozlov and A. I. Neishtadt, {\em Mathematical Aspects of Classical and Celestial Mechanics} 
(Springer, Berlin, 1997).} there are some crucial caveats regarding its implementation.  Even if one is lucky to 
identify enough first integrals, it doesn't necessarily follow that a ``closed form'' expression for the solution can 
be obtained. To begin with,  this would require the integrals to be not too complicated functions of the dynamical 
variables, allowing  explicit isolation of some coordinates at various stages; and even the minimalist choice of 
algebraic functions gives no guarantee.  Moreover, the operations involved in the method 
(function inversion,  integration, etc), can easily take one out of the set 
of ``elementary functions''. 

By the way, it should be emphasized that even for the two-body problem there is no 
such simple ``closed formula'' for the position of the bodies: it entails solving  
Kepler's equation, an implicit transcendental equation usually  solved by  a series expansion involving special 
functions.~\endnote{Ref. 1, pp. 85.} What is remarkable  about the two-body problem is that one 
can classify the possible orbits: they are the conic sections. 
This is in stark contrast to the three-body problem whose possible motions are {\em much} more complex 
and poorly  understood.

Near the end of the 19th century there was a growing suspicion that maybe there aren't enough
integrals for the three-body problem. Then, the  hope to find ``simple'' first integrals was dashed by an 
impossibility (or ``no-go'')  theorem due to Heinrich  Bruns  (1887) which states that no additional 
independent integrals exist which are algebraic functions of position and velocities 
(in Cartesian coordinates).~\endnote{The proof contained some mistakes. It 
was later corrected and extended by Poincar\'e and Painlev\'e, but still contained some errors which
were cleared only recently in E. Julliard-Tosel, ``Brun's Theorem: the Proof and Some
Generalizations'', Celest. Mechs., {\bf 76}, 241-281 (2000).}

Contrary to what is frequently asserted, this doesn't mean that the three-body problem is unsolvable, 
just that the {\em method} of quadratures won't work for it. By that time, solutions expressed  as 
an infinite {\em series} was increasingly gaining acceptance (after all, perturbative expansions 
had been used for quite some time) and there was great anticipation for the general {\em exact} series
solution to  the $N$-body problem. This optimism is quite evident in the announcement 
of the famous prize problem sponsored by  King Oscar II of Sweden in 1885:~\endnote{Quoted from
J. Barrow-Green, {\em Poincar\'e and the Three-Body Problem} (American Mathematical Society, 1997),
pp. 229.}
\begin{quote}
``A system being given of a number whatever of particles attracting one another mutually according to
Newton's law, it is proposed, on the assumption that there never takes place an impact of two particles to 
expand the coordinates of each particle in a series proceeding according to some known functions of time
and converging uniformly for any space of time.

It seems that this problem, the solution of which will considerably enlarge our knowledge with regard to the
system of the universe, might be solved by the analytical resources at our present disposition;...''
\end{quote}
Moreover,  it was widely believed that the old problem of the stability of the solar system 
(which goes back to Newton) would then be settled once and for all.

As is well known, Poincar\'e won the prize, not by actually solving the  problem, but due to the
many outstanding new ideas and methods he devised to attack it. One surprising result concerned 
the Lindstedt series, a perturbation  expansion widely used in celestial mechanics: he showed that 
while using a few terms of the series worked well,  the complete series was usually divergent!

 So, near the end of the 19th century a disturbing  collection of negative results began to accumulate, 
that brought celestial mechanics to a sort of crisis centered on the vexing  question of  what should count 
as a {\em bona fide} solution of the three-body problem. Some 30 years later, the Finnish astronomer 
Karl F. Sundman finally  solved the prize problem exactly as required but, alas,  it  didn't fulfill the 
expectations.

\renewcommand{\thesection}{\Roman{section}}
\section{Singularities and the total collapse theorem}

Sundman was after global solutions, i.e., valid for all time, as required by the prize problem. 
Thus, the first difficulty he had to overcome was that of avoiding  initial conditions that lead to a
{\em singularity}. At this point, it is important to recall that the EUT is a purely {\em local} result: it
only guarantees the existence of solutions for a small time interval  around the initial time $t_0$. 
Of course, one could then try to enlarge this existence interval by patching up: pick a time instant  $t_1$ 
near the boundary of that interval and use $({\bf x}(t_1), {\bf p}(t_1))$ as new 
initial conditions; apply the  EUT to enlarge the interval, and so on. If this can be continued forever, we
eventually obtain a global or {\em regular} solution, which is defined for all times past and future.
If, however,  there happens to exist a (future or past) time instant $t^*$  beyond which the solution {\em cannot} 
be extended, then we have what is called a {\em singular} solution.  As we take regular solutions for
granted, it is natural to ask: what is the nature of singularities? And how ``rare'' are they 
(in some sense)?

The first question above was dealt with by Paul Painlev\'e in 1897. As could be suspected, singularities 
might appear if particles get too close to each other so that the potential diverges and the equations 
of motion break down. In fact, Painlev\'e proved that a solution
$({\bf x}(t), {\bf p}(t))$ has a singularity at time $t^*$ if, and only if, 
$\rho(t) \equiv \min_{j\neq k} r_{jk}(t)$
goes to zero as $t$ approaches $t^*$ (where $r_{jk}={\|\R_k -\R_j\|}$).~\endnote{For details, see
 D. G. Saari, {\em Collisions, Rings and Other Newtonian
N-Body Problems} (American Mathematical Society, CBMS, Regional Conference Series in Mathematics, 
Number 104, Rhode Island, 2005).}

A {\em collision} is an obvious type of singularity, whose avoidance is clearly 
required in the prize problem.  In a collision at time $t^*$,  at least one pair of particles occupy
the same position. However, it is conceivable that  $\rho (t)$ could tend to zero without the occurrence of a 
collision:  as $t$ tends to  $t^*$,  the system  could experience
a sequence of  ``close approaches'', with particles almost colliding, but then separating, to come even closer later,
etc, in a wild oscillatory motion. This non-collisional type of singularity, dubbed a {\em pseudocollision},
is quite weird: in 1908 the Swedish astronomer Hugo von Zeipel proved that it necessarily follows that some particles of
the system go off to infinity in finite time.~\endnote{Though this is no contradiction with the model at hand, it is still remarkable that Newton's equations allow such bizarre solutions. In the known examples, use is made of
the  inexhaustible source of energy from Newtonian point-mass potential, see D. G. 
Saari and Z. Xia, ``Off to Infinity in Finite Time'', Notices of the AMS, {\bf 42} (5), 538-546 (1995).}  Painlev\'e proved that 
pseudocollisions do not occur in the three-body problem; unable to extend the proof he conjectured that they can appear 
for $N\geq 4$.~\endnote{This is  known as {\em Painlev\'e's Conjecture}, that was proved only
recently, for $N\geq 5$; the case $N=4$ it is still open. See F. Diacu, ``Painlev\'e's Conjecture'', 
The Mathematical Intelligencer, {\bf 15} (2), 6-12 (1993).}

Back to the three-body problem, and building on these results, Sundman first realized that 
{\em binary} collisions are only apparent singularities: they can be ``regularized'' by a suitable 
re-parametrization so that the motion is continued beyond them much as if the bodies  
bounced each other elastically.~\endnote{For a discussion, see D. G. Saari, ``A Visit to the Newtonian $N$-body Problem via
Elementary Complex Variables'', Am. Math. Monthly, {\bf 97} Feb., 105-119 (1990).} Unfortunately this trick doesn't  work for 
{\em ternary} collisions~\endnote{Incidentally, the same happens for ternary collisions 
of hard  ``billiard balls'': in contrast to  binary collisions, the conservation laws are not enough 
to unambiguously fix the posterior motion.} and therefore he looked for a criterion to avoid them.

Now, a ternary collision in the  three-body problem  is an instance  of a {\em total collapse}: 
all particles collide  at the same time at the same position. Sundman's  criterion for avoiding these
``catastrophes'' is the content of his {\em total collapse theorem}, namely:
if the system's total angular momentum ${\bf c}$ (a measure 
of the its global rotation) is non-zero, then collapse cannot happen (equivalently, if collapse 
occurs, then necessarily ${\bf c}= {\bf 0}$).~\endnote{The converse is not
true, except for the two-body problem. Examples are motions confined to a line, but then binary collisions
necessarily occur in the future or the past. Another (collision less) example is the recently discovered 
figure {\em eight} periodic solution of the three-body problem.}

The proof, valid for the general $N$-body problem,  is based on two relatively simple results of intrinsic
interest: the {\em   Lagrange-Jacobi's  identity} and {\em Sundman's inequality.} Both are expressed in terms
of the system's  {\em moment of inertia}: 
\begin{equation}
\label{inert1}
I \equiv \frac{1}{2}  \sum_{j=1}^N m_j \R_j^2,
\end{equation}
which is a measure of the spatial distribution of the masses.~\endnote{We follow the tradition
in celestial mechanics of putting a $1/2$ factor.}
As we are using the center-of-mass frame, a simple algebra shows that:
\begin{equation}
\label{inert2}
I =  
\frac{1}{2M}  \sum_{j < k}^N m_j m_k  r_{jk}^2, 
\end{equation} 
where $M$ is the total mass of the system. For future use, we note that Eq.~\eqref{inert2} implies that 
if total collapse happens at time $t^*$, then $I(t^*)=0$; thus, from Eq~\eqref{inert1}, it follows that 
the system collapses at the origin.

\begin{lemma}[Lagrange-Jacobi's identity]
Let $h$ be the total energy of the system, then:
\begin{equation}
\label{lag}
\ddot{I}= 2T-U =T+h=U+2h.
\end{equation}
\end{lemma}

{\bf Proof:}
The last two equalities in~\eqref{lag} follow from energy conservation. As for the first one, 
we just carry out the two derivatives using the chain-rule. We have:
\begin{equation}
\label{lag1}
\dot{I}= \sum_{j=1}^N m_j \R_j \cdot \V_j \Rightarrow \ddot{I} =\sum_{j=1}^N m_j \V_j^2 +
\sum_{j=1}^N m_j \R_j \cdot \ddot{\R}_j 
=2 T + \sum_{j=1}^N \R_j \cdot \nabla_{\R_j} U,
\end{equation}
where the last equality follows by plugging in Newton's equation \eqref{motion}. Now,  the potential energy is a
homogeneous function of degree $-1$, that is:
$\displaystyle U(\lambda{\bf x}) = \lambda^{-1} U({\bf x})$. Applying  Euler's theorem of advanced 
calculus~\endnote{See J. E. Marsden and A. J. Tromba, {\em Vector Calculus} (W H Freeman and Company,
San Francisco, 1981), 2nd. ed., pp. 148, ex. 21.} we then get:
\begin{equation}
\label{hom}
-U({\bf x}) = \sum_{j=1}^N \R_j \cdot \nabla_{\R_j} U ({\bf x}),
\end{equation}
which, when substituted in Eq.~\eqref{lag1}, gives the first identity in~\eqref{lag}.

\begin{lemma}[Sundman's inequality]

Consider the system's total angular momentum
$\displaystyle {\bf c} = \sum_{j=1}^N m_j \R_j \wedge \V_j$ and let  $c=\| {\bf c} \|$. Then,
\begin{equation}
\label{sund}
c^2 \leq 4 I (\ddot{I} - h).
\end{equation}
\end{lemma}

{\bf Proof:}
This is a nice  application of the Cauchy-Schwarz inequality from linear algebra: 
\begin{eqnarray}
\label{cau}
c \leq  \sum_{j=1}^N m_j \| \R_j \wedge \V_j\| \leq  \sum_{j=1}^N m_j \|\R_j\|  \|\V_j\| =
\sum_{j=1}^N (\sqrt{m_j} \|\R_j\|) (\sqrt{m_j} \|\V_j\|) & & \\
\leq \sqrt{\sum_{j=1}^N m_j \R_j^2} \sqrt{\sum_{j=1}^N m_j \V_j^2} = \sqrt{2I} \sqrt{2T}
= \sqrt{4I T}.& &
\end{eqnarray}
In sum, we have:
\begin{equation}
\label{bosta}
c^2 \leq 4 I T, 
\end{equation}
from which we get~\eqref{sund} by using  Lagrange-Jacobi's identity,~\eqref{lag}.

We can at last state and prove Sundman's theorem (already known to Weierstrass for the three-body problem).
\begin{theo}[The total collapse theorem]
If total collapse happens, then ${\bf c}= {\bf 0}$.~\endnote{Sundman
proved a stronger result: if  ${\bf c}\neq {\bf 0}$, then o $\max_{i\neq k} r_{ik} \geq D({\bf c})>0$, so that
the particles remain  strictly isolated from triple collisions.}
\end{theo}

 {\bf Proof:} Let $t^*$ be the time of total collapse, assumed positive without loss of generality.
As a collapse  is a multiple collision, we have:
\begin{equation}
\label{div}
\lim_{t \rightarrow t^*} U(t) = +\infty.
\end{equation}
Therefore, by Lagrange-Jacobi's identity we get:
\begin{equation}
\label{conv}
\lim_{t \rightarrow t^*} \ddot{I}(t) = +\infty.
\end{equation}
From~\eqref{conv} it follows that for all  $t$ in a neighborhood of  $t^*$, with $t < t^*$, we have  $\ddot{I}(t) >0$. As 
$I(t) > 0$ and recalling that $I(t^*)=0$, it
follows from  calculus that $I(t)$ is strictly decreasing function in this neighborhood. Hence,
for $t_1 \leq t \leq  t_2$, with $t_2 < t^*$, we have $-\dot{I}(t) >0$.

 Now, consider Sundman's inequality in the form:
\begin{equation}
\label{format}
\ddot{I}(t) \geq \frac{c^2}{4 I(t)} + h,
\end{equation}
for  $t_1 \leq t \leq  t_2$. Multiplying~\eqref{format} by $-\dot{I}(t)>0$, we get:
\begin{equation}
\label{format1}
-\dot{I} \ddot{I} \geq -\frac{c^2}{4}\frac{\dot{I}}{I} - h \dot{I},
\end{equation}
or better still:
\begin{equation}
\label{format2}
-\frac{1}{2} \frac{d}{dt} (\dot{I})^2 \geq -\frac{c^2}{4}\frac{d}{dt} \ln (I)- h \frac{d}{dt}I.
\end{equation}
Integrating both sides of~\eqref{format2}  from $t_1$ to  $t_2$, we get:
\begin{equation}
\label{format3}
-\frac{1}{2}[\dot{I}^2(t_2)-\dot{I}^2(t_1)] \geq \frac{c^2}{4}\ln[I(t_1)/I(t_2)]- h[I(t_2)-I(t_1)],
\end{equation}
or, regrouping,
\begin{equation}
\label{format4}
\frac{c^2}{4}\ln[I(t_1)/I(t_2)] 
\leq  h[I(t_2)-I(t_1)] +\frac{1}{2}[\dot{I}^2(t_1)-\dot{I}^2(t_2)].
\end{equation}
But, $I(t_2)-I(t_1) \leq I(t_2)$ and $\dot{I}^2(t_1)-\dot{I}^2(t_2) \leq \dot{I}^2(t_1)$, so from~\eqref{format4}
we get:
\begin{equation}
\label{format5}
\frac{c^2}{4} \leq \frac{h I(t_2)+\dot{I}^2(t_1)}{\ln[I(t_1)/I(t_2)]}.
\end{equation}
Finally, observe that the right-hand side of~\eqref{format5} goes to zero as $t_2$ tends to $t^*$ 
(for each fixed $t_1$). As $c$ is constant, we conclude that $c=0$, which completes the proof.
\medskip 

From Sundman's theorem it follows that any initial conditions leading to collapse must 
belong to the set defined by the equations ${\bf c}= {\bf 0}$. But these equations specify a lower dimensional hyper-surface in
phase-space, having therefore zero volume (much as a circle has zero area in
the plane). {\em If} we then agree that subsets of zero volume are ``rare'' or ``atypical'' (and in a sense,
negligible), one could say that  ``typical'' (or ``practically all'') solutions of the $N$-body problem are
collapse-free, which is a bit reassuring.

\renewcommand{\thesection}{\Roman{section}}
\section{Aftermath and conclusions}

In 1912 Sundman succeeded in finding the general exact solution of the three-body problem as 
an infinite series in powers of $t^{1/3}$, for all time $t$, and for all  initial conditions 
(except for the negligible set leading to ternary collisions).~\endnote{For a discussion, see  M. 
Henkel, ``Sur la solution de Sundman du probl\`eme des trois corps'',  Philosophia Scientiae, {\bf 5} (2), 161-184 (2001) and for details of the proof see C. Siegel and J. Moser, {\em Lectures in Celestial Mechanics},
(Springer-Verlag, NY, 1971). The generalization for $N \geq 4$ was obtained much later, see Q. Wang, ``The global solution
of the $N$-body problem'', Celest.  Mechs, {\bf 50}, 73-88 (1991).} He thus cracked  a problem that had resisted the attempts
of the greatest  mathematical-physicists of all times; so, even if a little too late, the Oscar definitely should 
go to Sundman. 

But there is an ironic twist to the story. It soon became clear that Sundman's solution gave no new
information about the general behavior of the system. Going from bad to worse, it was later 
shown that it is  also useless for numerical calculations due to  the incredibly slow rate of 
convergence of the series: it is estimated that of the order of $10^{8.000.000}$ terms would be needed to 
match the precision of astronomical measurements!~\endnote{D. Belorizky, ``Recherches sur l'application
pratique des solutions g\'en\'erales du probl\'eme des trois corps'', Journal des Observateurs,
{\bf 16} (7), 109-131 (1933).} Hence, though justly hailed  as an important conceptual achievement, 
his solution  didn't lay bare the secrets of the three-body problem, probably explaining
why it is hardly mentioned in the physics literature. 

Somewhat paradoxically,  Sundman's solution shows that  finding  an 
exact solution does not always improve our understanding of a problem. 
The positive side of all that  (and of the previous  impossibility results) was that
it  ultimately led to a momentous change from a predominantly {\em quantitative} to a new
{\em qualitative} approach in dynamics, a culmination of a long historical process.~\endnote{See  M. W. Hirsch, 
``The Dynamical System's Approach to Differential Equations'',  Bulltin of the AMS, {\bf 11} (1), (1984), section 6, and
also  A. Chenciner , De la M\'ecanique c\'eleste \'a la th\'eorie des
syst\'emes dynamiques, aller et retour, in actes de la conférence "Epistémologie des systèmes dynamiques", Paris Décembre 1999, to
appear.} Pioneered by Poincar\'e, Lyapunov and others, it  realizes that in dealing with such complicated  
dynamical systems as the $N$-body problem,  the focus should move from finding particular 
solutions to a study of {\em families} of them (and even of families of systems!). One should try to figure 
out, as an expert said,   ``most of the dynamics of most systems'',~\endnote{J.-C. Yoccoz, ``Recent developments in dynamics'',
in {\em Proceedings of the International Congress of Mathematicians}, {\bf 1} (Birkh\"auser, Basel, 1995) 
pp. 247-265.} that is, the  {\em typical} behavior and properties (for instance, but not exclusively, 
those valid except for a ``small'' set of initial conditions). Elsewhere, we
argued that a similar change was taking place, more or less at the same time, 
in  Ludwig Boltzmann's statistical approach to kinetic gas theory.~\endnote{See
S. B. Volchan, ``Probability as typicality'', Studies in History and Philosophy of Modern Physics, {\bf 38} (4), 
801-814 (2007).}

 Another  interesting aspect of this story is the emergence of the singularities. In spite of 
appearing in many branches of physics (e.g., in statistical mechanics, hydrodynamics, general relativity) 
it is surprising the lack of attention devoted to a unified understanding of  their role and meaning. An
investigation could  bring some additional understanding to the delicate question of  mathematical idealizations in
physical models. There are many interesting issues to examine: is a singularity a signal of the 
breakdown of physical laws; is it a mathematical way to describe some underlying peculiar (``non-smooth'') 
phenomenon; or is it  just an artifact arising from a simplified model? 
In the case of the $N$-body problem it could be argued that the last case applies, as the use of  
point masses, with their infinite source  of potential energy, is physically untenable. However, this model
is pretty robust in capturing some essential features of real systems. In this sense it would be
nice to have at least a  proof that singularities are rare. For the moment, it is known that the set of initial 
data leading to collisions  of any kind has zero volume, for all $N$. As for singularities of any kind,
the same is true in the  $N=4$ case, while it is an open problem for $N\geq 5$. In other words, one doesn't even know whether or 
not the general $N$-body problem has global solutions for ``most'' initial conditions!

 It is said that Newton complained of  a headache when he tackled the three-body problem in the guise of the 
Sun-Earth-Moon system. Today, with more than 300 years of hindsight and  despite the great advances in 
non-linear dynamics, numerical analysis and computer simulations, even the experts admit that  
``the three-body problem is as enigmatic as ever''.~\endnote{C. D. Murray and S. F. Dermott, 
{\em Solar System Dynamics} (Cambridge University Press, NY, 2005), p. 63.}  The headache  
continues and it seems it will persist for quite some time.

\bigskip

\begingroup
\setlength{\parindent}{0pt}\setlength{\parskip}{2ex}
\renewcommand{\enotesize}{\normalsize}
\theendnotes\endgroup

\bigskip

\noindent{\bf S\'ergio B. Volchan} 

\noindent{\em
Departamento de Matem\'atica, Pontif\'{\i}cia Universidade Cat\'olica do 
Rio de Janeiro, Rua Marqu\^es de S\~ao Vicente 225,
G\'avea, 22453-900  Rio de Janeiro, Brasil

\noindent volchan@mat.puc-rio.br}

\end{document}